\documentclass[12pt]{article}
\usepackage{epsfig}
\usepackage{graphicx}
\usepackage{amsmath,lscape,sidecap}
\usepackage{booktabs}
\usepackage{amssymb}

\def\beq{\begin{equation}}
\def\eeq#1{\label{#1}\end{equation}}
\def\eeqn{\end{equation}}

\def\beqa{\begin{eqnarray}}
\def\eeqa#1{\label{#1}\end{eqnarray}}
\def\eeqan{\end{eqnarray}}

\let\bar=\overbar

\def\Dslash{\not{\hbox{\kern-4pt $D$}}}
\def\dslash{\not{\hbox{\kern-2pt $\del$}}}

\def\msb{{\bar{\ssstyle M \kern -1pt S}}}

\usepackage{fancyhdr,graphicx}
\def\Title#1{\begin{center} {\Large {\bf #1} } \end{center}}

\begin{document}

\Title{Core-collapse supernova matter: light clusters, pasta phase and phase transitions}

\bigskip\bigskip


\begin{raggedright}

{\it 
Helena Pais$^1$, Constan\c ca Provid\^encia$^1$, William G. Newton$^2$ and Jirina R. Stone$^{3,4}$\\
\bigskip
  $^1$Centro de F\'isica Computacional, Department of Physics, University of Coimbra\\ 
 P-3004-516 Coimbra -- Portugal\\
 \bigskip
 $^2$Department of Physics and Astronomy, Texas A\&M University, Commerce, TX 75429-3011, USA \\
\bigskip
 $^3$Department of Physics, University of Oxford, Oxford OX1 3PU,  United Kingdom \\
 \bigskip
 $^4$Department of Physics and Astronomy, University of Tennessee, Knoxville, TN 37996, USA \\
}
\end{raggedright}

\section{Introduction}

The complex structure of nuclear matter in the density region approaching $\rho_s \sim 0.16$ fm$^{-3}$  (central density of heavy nuclei) at finite temperature ($T < 20$ MeV) critically affects many astrophysical and nuclear physics phenomena. At low densities, the frustrated system called the ``pasta phase'', caused by the competition between the Coulomb interaction and the strong force, appears, and is constituted by different geometrical configurations, as the density increases \cite{Ravenhall-83, Horowitz-05, Maruyama-05, Watanabe-05a, Sonoda, Pais-12, Grill-14}.

The main interest in the pasta phase in core-collapse supernovae (CCSN) is that the neutrino opacity, which plays the main role in the development of a shock wave during the supernova collapse, is affected by its presence \cite{neutrinos}.
At very low densities (up to 0.001 times the saturation density), light nuclei (deuterons, tritons, helions, $\alpha$-particles) can appear, and, like the pasta phase, can modify the neutrino transport, which will have consequences in the cooling of the proto-neutron star \cite{Lattimer-91, Shen-98, Avancini-10Cl, Avancini-12Cl}. 

In this work, we use different methods to study the pasta phases of CCSN matter: a non-relativistic method, the 3DHFEOS method, where a 3D Skyrme-Hartree-Fock approximation is used, and relativistic mean field (RMF) methods, the Thomas-Fermi (TF) and the Coexisting Phases (CP) approximations, and a thermodynamical spinodal (TS) calculation.  

\section{The 3DHFEOS method}

We use a 3D Hartree-Fock approximation with a phenomenological Skyrme model (3DHFEOS) for the nuclear force \cite{Newton-09, Pais-12, Pais-14TN}. In the calculation, it is assumed that, at a given density and temperature, matter is arranged in a periodic structure throughout a sufficiently large region of space for a unit cell to be identified, in which the microscopic and bulk properties of the matter are calculated. The calculation is performed in cubic cells with periodic boundary conditions and assuming reflection symmetry across the three Cartesian axes. The required reflection symmetry allows us to obtain solutions only in one octant of the unit cell, which reduces significantly the computer time. The only effect of confining ourselves to $1/8$ of the cell is that we can only consider triaxial shapes. It is expected that the absolute minimum of the free energy of a cell containing $A$ nucleons is not going to be particularly pronounced and there will be a host of local minima separated by relatively small energy differences. In order to systematically survey the ``shape space" of all nuclear configurations of interest, the quadrupole moment of the neutron density distributions has been parametrized, and those parameters constrained. It is expected that the proton distribution follows closely that of the neutrons. The minimum of the free energy in a cell at a given particle number density, temperature and a proton fraction is sought as a function of 3 free parameters, the number of particles in the cell (determining the cell size) and the parameters of the quadrupole moment of the neutron distribution $\beta, \gamma$. 
We selected four different interactions, SkM* \cite{Bartel-82}, SLy4 \cite{Chabanat-98}, NRAPR \cite{Steiner-05} and SQMC700 \cite{Guichon-06}, based on their overall performance in modeling of a wide variety of nuclear matter properties \cite{Dutra-12}. 

\section{The RMF approximation}

We use relativistic mean field nuclear parametrizations with constant couplings and nonlinear terms, FSU \cite{Todd-Rutel-05} and IUFSU \cite{Fattoyev-10}, and with density-dependent couplings, TW \cite{Typel-99}. The FSU model is too soft and does not describe a 2 $M_\odot$ neutron star, however it is expected to describe well the crust \cite{Grill-14}.

The equations are derived from the Lagrangian density, where we consider a system of nucleons and electrons, and is given by
\begin{equation*}
    \mathcal{L}=\sum_{i=p,n}{\mathcal{L}_{i}} +\mathcal{L}_{e } 
    + \mathcal{L}_{{\sigma }}+ \mathcal{L}_{{\omega }} + 
\mathcal{L}_{{\rho }}+{\mathcal{L}}_{\gamma } + {\mathcal{L}}_{nl},
\end{equation*}

where the nucleon Lagrangian reads

\begin{equation}
\mathcal{L}_{i}=\bar{\psi}_{i}\left[ \gamma _{\mu }iD^{\mu }-M^{*}\right]
\psi _{i}  \label{lagnucl},
\end{equation}

with
\begin{eqnarray}
iD^{\mu } &=&i\partial ^{\mu }-\Gamma_{\omega}\omega^{\mu }-\frac{\Gamma_{\rho }}{2}{\boldsymbol{\tau}}%
\cdot \boldsymbol{\rho}^{\mu } - e \frac{1+\tau_3}{2}A^{\mu}, \label{Dmu} \\
M^{*} &=&M-\Gamma_{\sigma}\sigma
\end{eqnarray}
and the electron Lagrangian is given by
\begin{equation}
\mathcal{L}_e=\bar \psi_e\left[\gamma_\mu\left(i\partial^{\mu} + e A^{\mu}\right)
-m_e\right]\psi_e.
\label{lage}
\end{equation}
The meson and electromagnetic Lagrangian densities are
\begin{eqnarray*}
\mathcal{L}_{{\sigma }} &=&\frac{1}{2}\left( \partial _{\mu }\sigma \partial %
^{\mu }\sigma -m_{\sigma}^{2}\sigma ^{2}\right) \\ 
\mathcal{L}_{{\omega }} &=&\frac{1}{2} \left(-\frac{1}{2} \Omega _{\mu \nu }
\Omega ^{\mu \nu }+ m_{\omega}^{2}\omega_{\mu }\omega^{\mu } \right) \\
\mathcal{L}_{{\rho }} &=&\frac{1}{2} \left(-\frac{1}{2}
\mathbf{R}_{\mu \nu }\cdot \mathbf{R}^{\mu
\nu }+ m_{\rho }^{2}\boldsymbol{\rho}_{\mu }\cdot \boldsymbol{\rho}^{\mu } \right)\\
\mathcal{L}_{{\gamma }} &=&-\frac{1}{4}F _{\mu \nu }F^{\mu
  \nu }\\
\mathcal{L}_{{nl}} &=&-\frac{1}{3!}\kappa \sigma ^{3}-\frac{1}{4!}%
\lambda \sigma ^{4}+\frac{1}{4!}\xi \Gamma_{\omega}^{4}(\omega_{\mu}\omega^{\mu })^{2} 
+\Lambda_\omega \Gamma_\omega^2 \Gamma_\rho^2 \omega_{\nu
}\omega^{\nu } \boldsymbol{\rho}_{\mu }\cdot
\boldsymbol{\rho}^{\mu }
\end{eqnarray*}
where $\Omega _{\mu \nu }=\partial _{\mu }\omega_{\nu }-\partial
_{\nu }\omega_{\mu }$, $\mathbf{R}_{\mu \nu }=\partial _{\mu
}\boldsymbol{\rho}_{\nu }-\partial _{\nu }\boldsymbol{\rho}_{\mu
}-\Gamma_{\rho }(\boldsymbol{\rho}_{\mu }\times
\boldsymbol{\rho}_{\nu })$ and $F_{\mu \nu }=\partial _{\mu
}A_{\nu }-\partial _{\nu }A_{\mu }$. 

\subsection*{Light clusters}

To add light clusters ($d\equiv^2$H, $t\equiv^3$H, $\alpha\equiv^4$He, $h\equiv^3$He) to our system, we need to increase its degrees of freedom. The Lagrangian density becomes
\begin{equation}
    \mathcal{L}=\sum_{i=p,n,t,h}{\mathcal{L}_{i}}+\mathcal{L}_{\alpha} + \mathcal{L}_d+\mathcal{L}_{e } 
    + \mathcal{L}_{{\sigma }}+ \mathcal{L}_{{\omega }} + 
\mathcal{L}_{{\rho }}+{\mathcal{L}}_{\gamma } + {\mathcal{L}}_{nl}.
\end{equation}
 
 The $\alpha$ particles and the deuterons are described as in \cite{Typel-10}:
\begin{eqnarray}
\mathcal{L}_{\alpha }&=&\frac{1}{2} (i D^{\mu}_{\alpha} \phi_{\alpha})^*
(i D_{\mu \alpha} \phi_\alpha)-\frac{1}{2}\phi_\alpha^* \left(M_{\alpha}^*\right)^2
\phi_{\alpha},\\
\mathcal{L}_{d}&=&\frac{1}{4} (i D^{\mu}_{d} \phi^{\nu}_{d}-
i D^{\nu}_{d} \phi^{\mu}_{d})^*
(i D_{d\mu} \phi_{d\nu}-i D_{d\nu} \phi_{d\mu})-\frac{1}{2}\phi^{\mu *}_{d} \left(M_{d}^*\right)^2 \phi_{d\mu}.
\end{eqnarray}
 
The effective masses of the clusters, $M^*_j, j=d,t,h,\alpha$  and their binding energies are given in \cite{Avancini-12Cl}.

\section{The Thomas-Fermi and the Coexisting Phases approximations}
 
We use the Thomas-Fermi approximation to describe the non-uniform $npe$ matter inside the Wigner-Seitz unit cell,  that is taken to be a sphere, a cylinder or a slab in three, two, and one dimensions  \cite{Avancini-1012, Grill-14}. In this approximation, $npe$ matter is assumed locally homogeneous and at each point
its density is determined by the corresponding local Fermi momenta. In 3D we consider spherical symmetry and in 2D we assume axial symmetry around the z axis. The fields are assumed to vary slowly so that baryons can be treated as moving in locally constant fields at each point \cite{Maruyama-05}. In this approximation, the surface effects are treated self-consistently.
 
In the Coexisting Phases (CP) method, matter is organized into separated regions of higher and lower
density, the higher being the pasta phases, and the lower a background nucleon gas. The interface between these regions is sharp, and it is taken into account by a surface term and a Coulomb one in the energy density \cite{Avancini-12Cl}. This method is much faster than the TF one, and the Gibbs equilibrium conditions are used to get the lowest energy state. 

\section{The thermodynamical spinodal calculation}

Matter is stable with respect to fluctuations in density and composition, under conditions of constant volume and temperature, when the free energy density $\cal F$ is a convex function of the proton and neutron densities \cite{Prov-06}. These densities are associated with the chemical potentials $\mu_n = \frac{\partial \cal F}{\partial \rho_n}$ and $\mu_p = \frac{\partial \cal F}{\partial \rho_p}$ 
and the free energy curvature is given by
$$
\cal C = \left(\begin{array}{c c}
\dfrac{\partial^2 \cal F}{\partial^2 \rho_p} & \dfrac{\partial^2 \cal F}{\partial \rho_p \partial \rho_n}\\
\dfrac{\partial^2 \cal F}{\partial \rho_n \partial \rho_p} & \dfrac{\partial^2 \cal F}{\partial^2 \rho_n}
\end{array}  \right) = \left(\begin{array}{c c}
\dfrac{\partial \mu_p}{\partial \rho_p} & \dfrac{\partial \mu_n}{\partial \rho_p}\\
\dfrac{\partial \mu_p}{\partial \rho_n} & \dfrac{\partial \mu_n}{\partial \rho_n}
\end{array}  \right)
$$
The eigenvalues of this matrix are given by $$\lambda_{\pm}=\frac{1}{2}(\rm{Tr}(\cal C) \pm \sqrt{\rm{Tr}(\cal C)^{\rm{2}}-\rm{4}\rm{Det}(\cal C)}$$ and the eigenvectors by $$ \frac{\delta \rho_p^{\pm}}{\delta \rho_n^{\pm}} = \dfrac{\lambda_{\pm} - \frac{\partial \mu_n}{\partial \rho_n}}{\frac{\partial \mu_p}{\partial \rho_n}}.$$

The thermodynamical spinodal region is then defined as the surface in ($\rho, y_p, T$) space where the curvature of $\cal F$ is $< 0$. The intersection with the Equation of State gives the transition density to uniform matter.

\section{Results}

We show in Table \ref{tab1} the nuclear matter properties for the four Skyrme models, NRAPR, SQMC700, SkM$^*$ and SLy4, and for the three RMF parametrizations, FSU, IUFSU and TW, we use in this study.

\begin{table}[!htbp]
  \centering
 \begin{tabular}{cccccc}
    \hline
    \hline
    	Model & $\rho_0$ & $B/A$ & $K$ & $E_{sym}$ & $L$   \\
    \hline
    	NRAPR & 0.16 & -15.85 & 226 & 33 & 60 \\
	SQMC700 & 0.17 & -15.49 & 222 & 33 & 59  \\
	SkM* & 0.16 & -15.77 & 217 & 30 & 46 \\
	SLy4 & 0.16 & -15.97 & 230 & 32 & 46 \\
	FSU & 0.15 & -16.3 & 230 & 33 & 60.5 \\
	IUFSU & 0.155 & -16.40 & 231 & 31 & 47 \\
	TW & 0.15 & -16.3 & 240 & 33 & 55 \\
	\hline   
    \hline
  \end{tabular}\medskip{}
    \caption{Nuclear matter properties at saturation density $\rho_0$ (energy per particle $B/A$, incompressibility $K$, symmetry energy $E_{sym}$ and symmetry energy slope $L$) for the models studied. All the quantities are in MeV, except for $\rho_0$, given in fm$^{\rm -3}$.}
 \label{tab1}
\end{table} 

\begin{figure}
\begin{tabular}{c}
\includegraphics[width=0.7\textwidth]{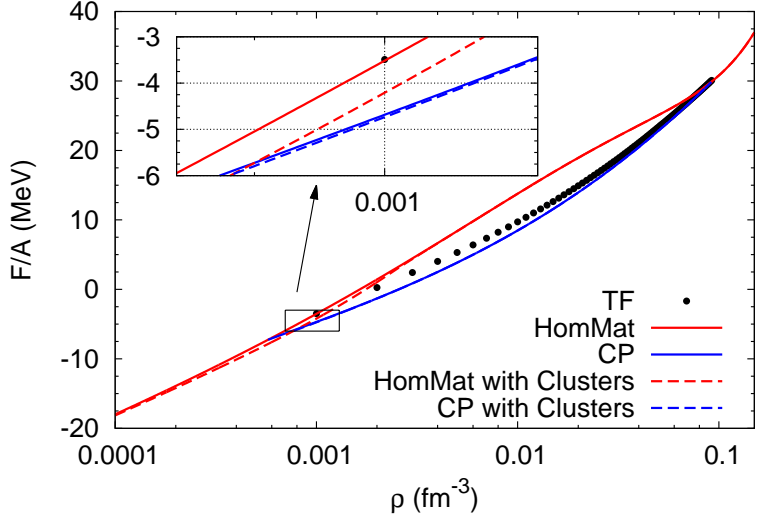} \\
\end{tabular}
\caption{Free energy per baryon as a function of the density for the FSU interaction, $T = 4$ MeV and $y_p=0.3$. Results with and without pasta, and including or not clusters, are shown. The effect of these aggregates are only seen for very small densities (inset panel).} 
\label{fig1}
\end{figure}

In Figure \ref{fig1}, we show the free energy per particle as a function of the density, for the FSU interaction, $T = 4$ MeV and $y_p= \rho_p/\rho_n = 0.3$. We show the results for homogeneous matter (red) and CP (blue) calculation with (dashed) and without (solid line) clusters. The TF results are shown with points. We can observe that $F$ is lowered when pasta is present, making these states the most stable ones. The light clusters are only present for very small densities, and will start melting for $\rho\gtrsim 0.001$ fm$^{-3}$, and their presence lowers the free energy.

Phase transitions can be classified as first or second (or higher) order (see e.g. \cite{Landau-80}). 
 For a first-order phase transition, at least one of the first derivatives of the free energy with respect to one of its variables is discontinuous: 
$S=-\frac{\partial F}{\partial T}|_{N,V,...}$
$P=-\frac{\partial F}{\partial V}|_{N,T,...}$.  
 This discontinuity produces a divergence in the higher derivatives like the specific heat $C_V = T\frac{\partial S}{\partial T}|_{V} = - T\frac{\partial^2 F}{\partial^2 T}|_{V}$, or the incompressibility $K(\rho_0)=9\rho_0^2\frac{\partial^2 E_{SNM}(\rho)}{\partial^2\rho}|_{\rho=\rho_0}$, where $E_{SNM}$ is the energy per particle of symmetric nuclear matter. 
For a phase transition of second (or $n$th order), the first derivatives of the free energy are continuous; however, second (or $n$th order) derivatives, like the specific heat or the susceptibility, are discontinuous or divergent. 

\begin{figure}
\begin{tabular}{c}
\includegraphics[width=1.\textwidth]{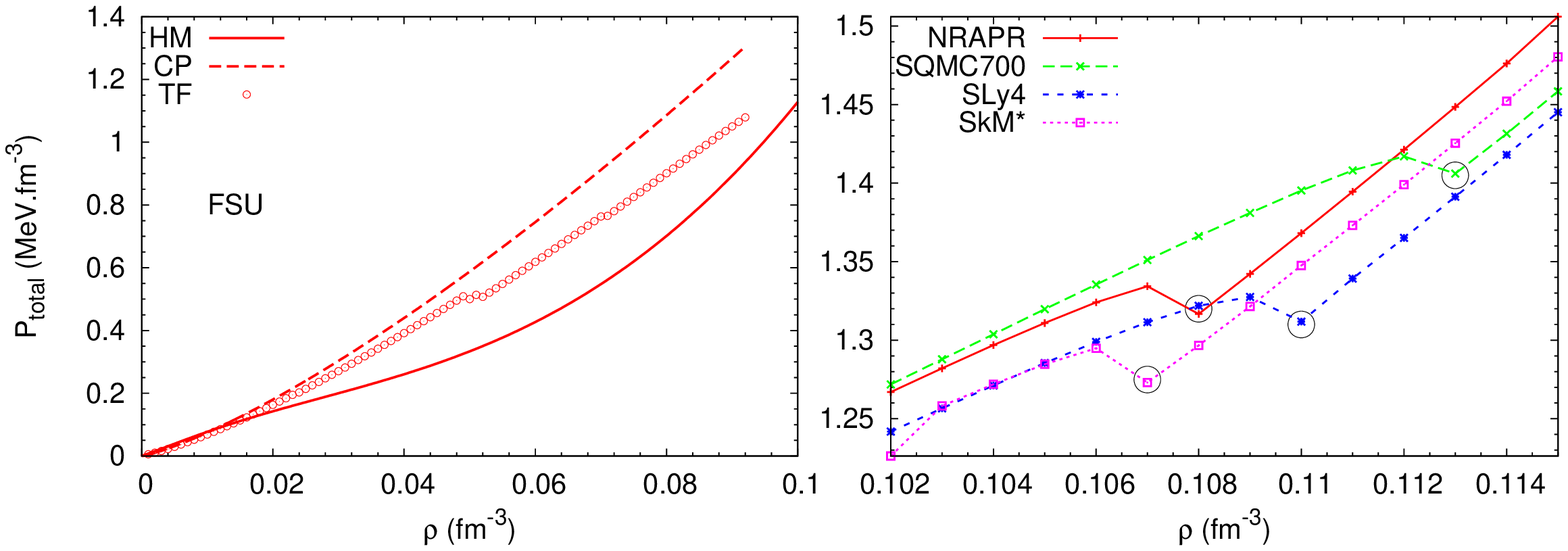} \\
\end{tabular}
\caption{Total pressure as a function of the density for the FSU (left) and Skyrme (right) parametrizations, $T = 4$ MeV and $y_p=0.3$. Results for the Thomas-Fermi (points), Coexisting Phases (dashed) and homogeneous matter (solid line) calculations are shown in the left panel. In the right panel, the transitions to homogeneous matter are circled. (Right plot taken from \cite{Pais-14TN}.)} 
\label{fig2}
\end{figure}

In Figure \ref{fig2}, we plot the total pressure as a function of the density. We can observe that the homogeneous matter pressure curvature is partially removed with the TF and CP calculations, though CP gives a larger correction, which may not be so realistic. The intermediate transitions (between shapes) are given by a discontinuity in the total pressure for the TF calculation. In the 3DHFEOS approximation, a discontinuity is shown for the transition to homogeneous matter, and is circled in the right plot.

\begin{figure}
\begin{tabular}{c}
\includegraphics[width=1.\textwidth]{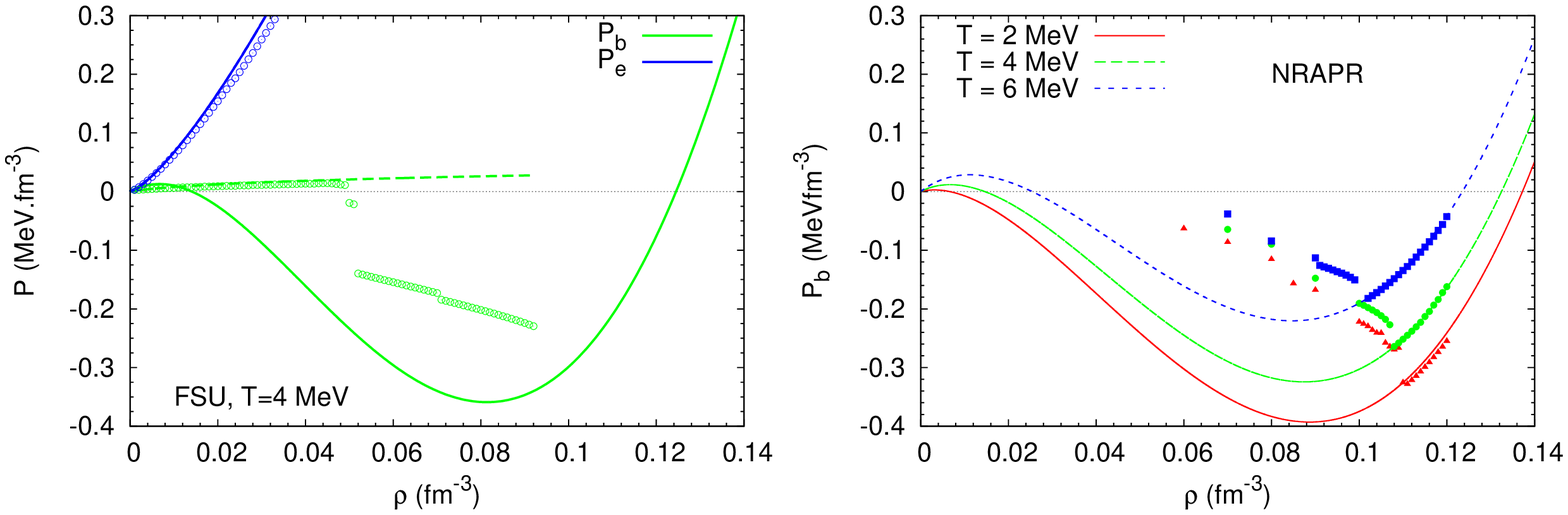} \\
\end{tabular}
\caption{Left: Baryonic (yellow) and electronic (blue) pressure components as a function of the density for the FSU interaction for $T = 4$ MeV and $y_p=0.3$. Results for the Thomas-Fermi (points), Coexisting Phases (dashed) and homogeneous matter (solid line) calculations are shown. 
Right: (Plot taken from \cite{Pais-14TN}) Baryonic pressure for the Skyrme interactions in the 3DHFEOS (points) and uniform matter (lines) calculations, for $T = 2, 4, 6$ MeV, and $y_p=0.3$.} 
\label{fig3}
\end{figure}

In Figure \ref{fig3}, we plot the baryonic pressure as a function of the density. Like in Fig.~\ref{fig2}, the homogeneous matter baryonic pressure curvature is partially removed within all methods, TF, CP, and 3DHFEOS, but the baryonic pressure still exhibits a region with a negative incompressibility. The electrons are responsible for making the total pressure positive.

\begin{figure}
  \resizebox{\linewidth}{!}{
  \begin{tabular}{c c c c c c c}
    \hline
    \hline
    \multicolumn{7}{c}{ $\rho$ [fm$^{\rm -3}$]}     \\
    \hline
    0.020 & 0.032(2) & 0.042(2) & 0.052(2) & 0.090(2) & 0.108(2) & 0.114(1)\\
\hline    
 \includegraphics[width=0.07\textwidth]{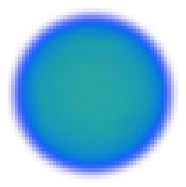} &
\includegraphics[width=0.07\textwidth]{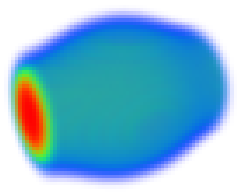} &
 \includegraphics[width=0.07\textwidth]{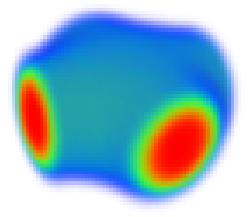} &
 \includegraphics[width=0.07\textwidth]{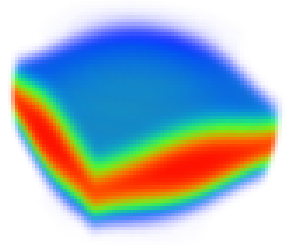} &
 \includegraphics[width=0.07\textwidth]{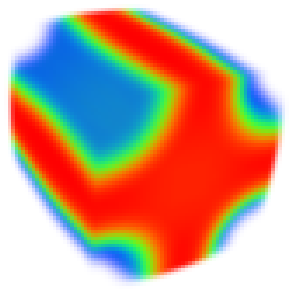} &
 \includegraphics[width=0.07\textwidth]{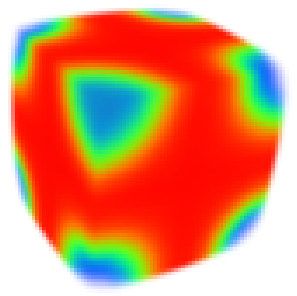} &
 \includegraphics[width=0.07\textwidth]{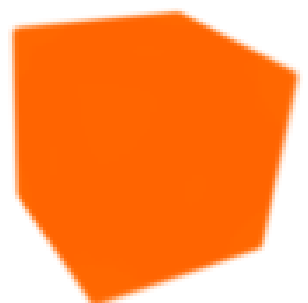} \\
     \hline
 \includegraphics[width=0.07\textwidth]{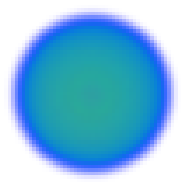} &
\includegraphics[width=0.07\textwidth]{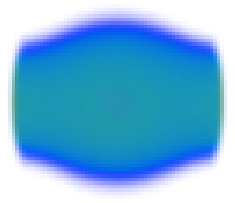} &
 \includegraphics[width=0.07\textwidth]{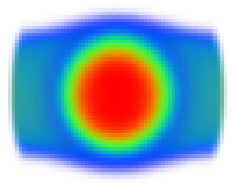} &
 \includegraphics[width=0.07\textwidth]{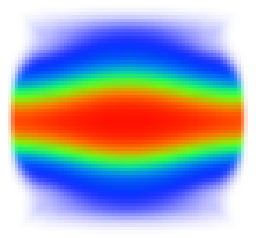} &
 \includegraphics[width=0.07\textwidth]{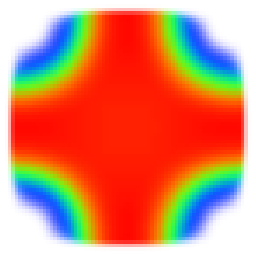} &
 \includegraphics[width=0.07\textwidth]{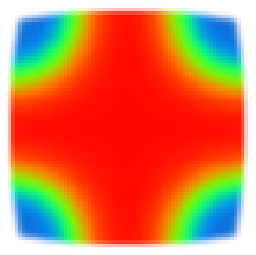} &
 \includegraphics[width=0.07\textwidth]{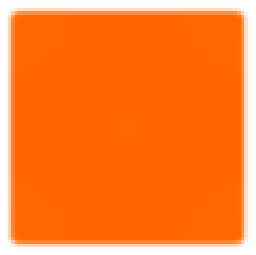} \\
     \hline
 \includegraphics[width=0.07\textwidth]{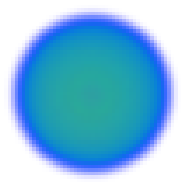} &
\includegraphics[width=0.07\textwidth]{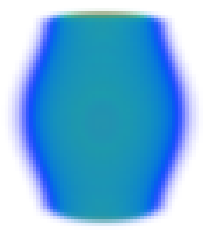} &
 \includegraphics[width=0.07\textwidth]{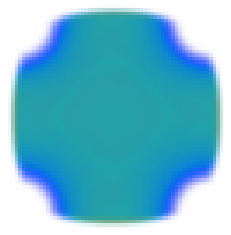} &
 \includegraphics[width=0.07\textwidth]{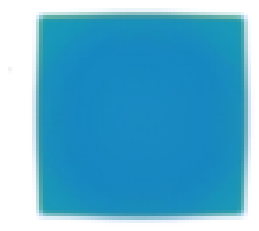} &
 \includegraphics[width=0.07\textwidth]{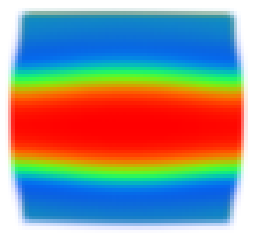} &
 \includegraphics[width=0.07\textwidth]{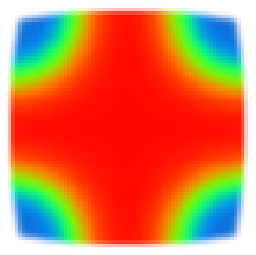} &
 \includegraphics[width=0.07\textwidth]{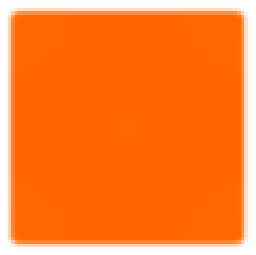} \\
     \hline
 \includegraphics[width=0.07\textwidth]{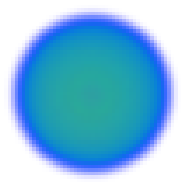} &
\includegraphics[width=0.07\textwidth]{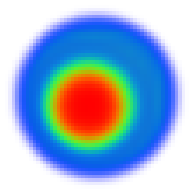} &
 \includegraphics[width=0.07\textwidth]{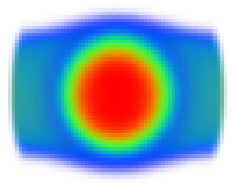} &
 \includegraphics[width=0.07\textwidth]{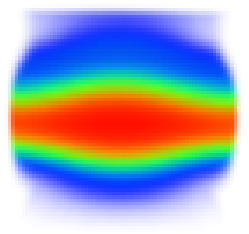} &
 \includegraphics[width=0.07\textwidth]{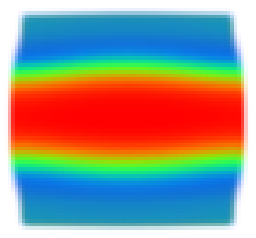} &
 \includegraphics[width=0.07\textwidth]{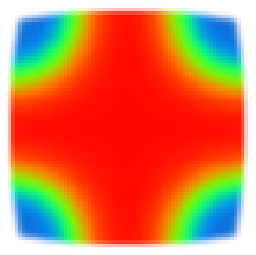} &
 \includegraphics[width=0.07\textwidth]{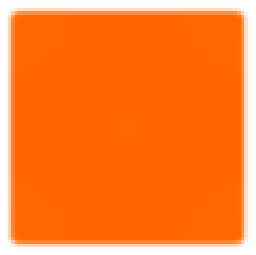} \\
     \hline
     \hline
  \end{tabular}}
\caption{(Figure taken from \cite{Pais-12}) The neutron density distribution for the SQMC700 model, and $T=2$ MeV and $y_p=0.3$, is shown at the density corresponding to the onset of each phase, known with the uncertainty given in brackets.  Blue (red) color indicates the bottom (top) of the density scale: 0.001 (dark blue) - 0.02475 (light blue) - 0.0485 (green) - 0.07225 (light orange) - 0.095 (red) fm$^{\rm -3}$.}
\label{fig4}
\end{figure}

Let us now discuss the evolution of the geometrical pasta shapes with the density. In Figure \ref{fig4}, we show the neutron density distribution for the SQMC700 interaction and $T = 2$ MeV and $y_p = 0.3$ at the onset density for each shape we found. The pasta configurations shown here appear for all the Skyrme models, but the threshold density changes somewhat. At 0.042 fm$^{-3}$ a new shape, cross-rod, normally not considered in pasta calculations, appears. This was due to the freedom used to find the equilibrium configuration. 

\begin{figure}
   \begin{tabular}{c}
\includegraphics[width=1.\textwidth]{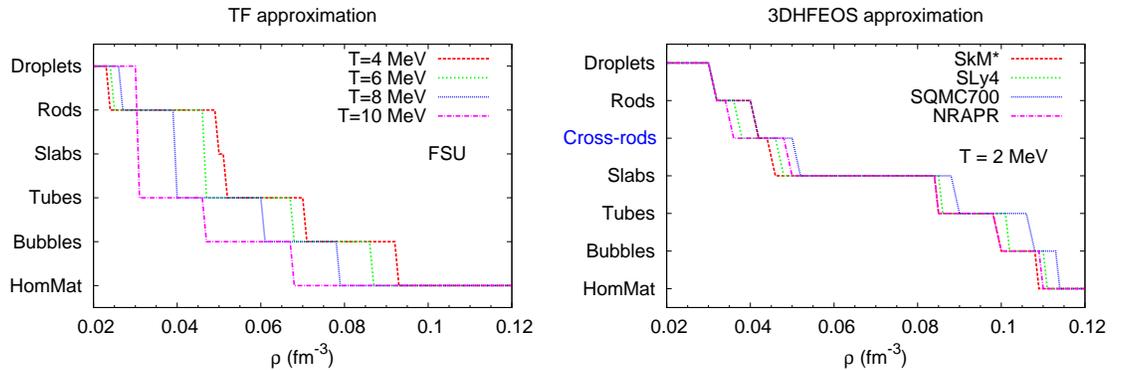}\\
   \end{tabular}
\caption{Transition densities between pasta formations. Left: FSU and $T=4-10$ MeV. Right (data taken from \cite{Pais-12}): Skyrme interactions and $T=2$ MeV.}   
\label{fig5}
\end{figure}

\begin{figure}
   \begin{tabular}{c}
\includegraphics[width=0.7\textwidth]{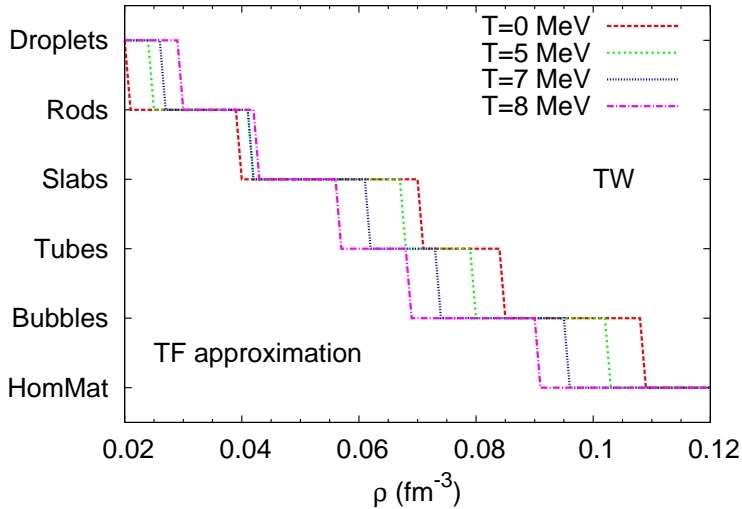}\\
   \end{tabular}
\caption{Transition densities between pasta formations for TW and $T=0-8$ MeV (data taken from \cite{Avancini-1012}).}   
\label{fig6}
\end{figure}

In Figure \ref{fig5}, we show the transition densities between the pasta configurations for the FSU interaction (left) and the Skyrme models (right). The slabs are omitted (except for $T = 4$ MeV) for the FSU parametrization and occupy the widest density range in Skyrme models considered. However, the slab geometry is present in other RMF parametrizations, as we can see in Figure \ref{fig6}. In FSU, the difference in the free energy between slabs and tubes  is less than $ 10^{-3}$ MeV. Therefore, the stable geometries depend on  the parametrizations considered, and the properties that influence them should be investigated. We can also observe from the left panel that the density range of each shape decreases with increasing temperature.%
Looking at the right panel of Fig.~\ref{fig5}, and as already seen in Fig.~\ref{fig4}, one more shape, cross-rods, is obtained with 3DHFEOS calculations. Also, the shapes and their range of densities are almost not affected by the choice of the Skyrme parametrizations, and this is probably due to the fact that we are using a number of Skyrme models that give very close predictions for nuclear matter properties (as seen in Table~\ref{tab1}).

\begin{figure}
   \begin{tabular}{c}
\includegraphics[width=1.1\textwidth]{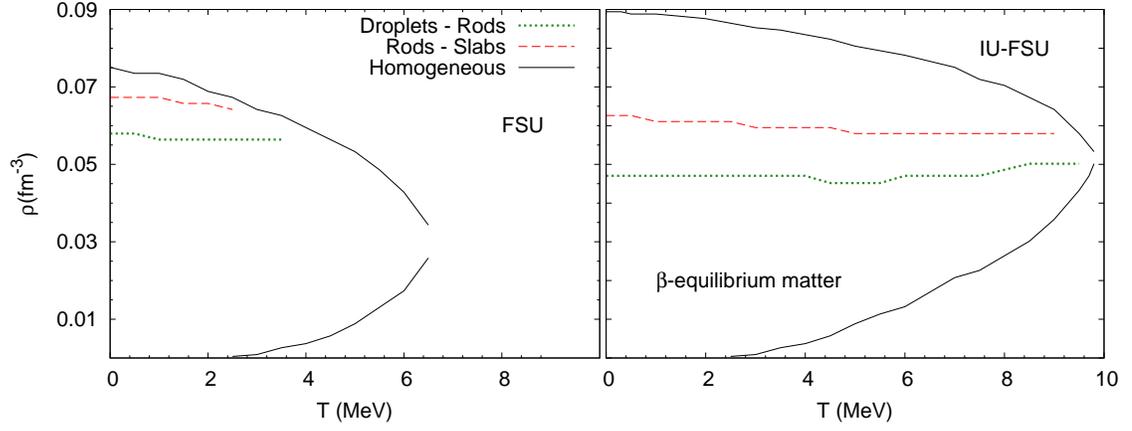}\\
   \end{tabular}
\caption{Transition densities between pasta formations for FSU (left) and IUFSU (right, plot taken from \cite{Grill-14}) as a function of $T$ for $\beta-$equilibrium matter.}   
\label{fig7}
\end{figure}

We have discussed so far the pasta configurations and their range of densities for a fixed proton fraction. Are there any differences if we consider $\beta-$equilibrium matter? Let us observe Figure \ref{fig7}, where the transition densities are plotted as a function of the temperature for FSU and IUFSU and $\beta-$equilibrium matter. We can see that the droplet-rod and rod-slab transition densities depend weakly on the temperature, and the melting $T$ is different for all shapes. IUFSU has pasta up to $T \sim 10$ MeV, but for other parametrizations pasta melts at $T \sim 4-6$ MeV. In comparison with $y_p=0.3$ matter, $\beta$-equilibrium matter only shows three stable configurations. FSU pasta melting $T$ is lower ($T \sim 6$ MeV) in this case than in fixed $y_p$ matter (see Fig.~\ref{fig6}). 

\begin{figure}
   \begin{tabular}{c}
\includegraphics[width=0.7\textwidth]{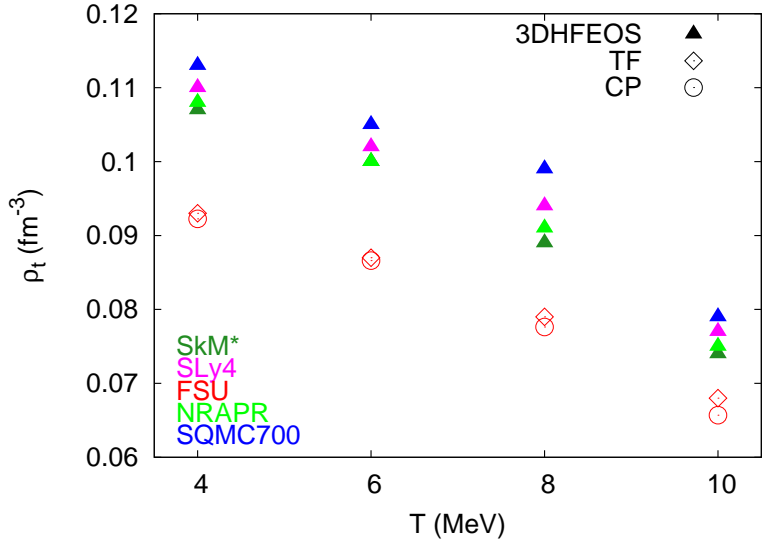}\\
   \end{tabular}
\caption{Transition densities to uniform matter. For more explanations, see text.  (3DHFEOS data taken from \cite{Pais-12}.)}   
\label{fig8}
\end{figure}

\begin{figure}
   \begin{tabular}{c}
\includegraphics[width=1.2\textwidth]{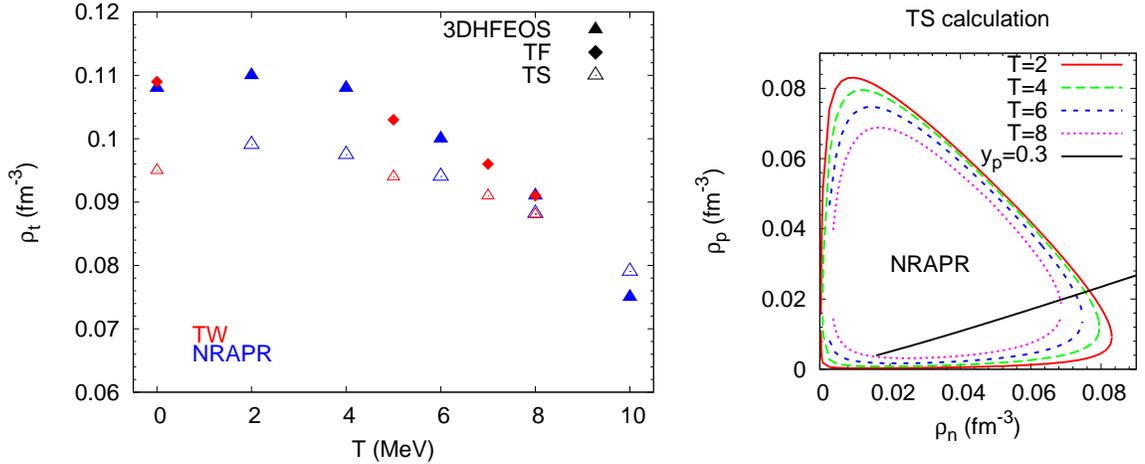} 
   \end{tabular}
\caption{Left: Transition densities to uniform matter. Right: TS calculation. For more explanations, see text.  (Left plot: NRAPR data taken from \cite{Pais-14TN} and TW data taken from \cite{Avancini-1012}. Right plot taken from \cite{Pais-14TN}.)}   
\label{fig9}
\end{figure}

Finally, in Figs.~\ref{fig8} and \ref{fig9} we plot the transition densities to uniform matter. In Fig.~\ref{fig8}, we compare the 3DHFEOS calculation with the four Skyrme interactions, with the CP and TF calculations for the FSU model. We can observe that the TF and CP calculations are almost coincident. The difference to the 3DHFEOS calculation is $\sim 0.015$ fm$^{-3}$ and decreases with increasing $T$.
In the left panel of Fig.~\ref{fig9}, we compare the results obtained with TF and TS for the TW model with the results obtained with the 3DHFEOS and TS for the NRAPR interaction. In the right panel, we demonstrate how to obtain the transition density to uniform matter within the thermodynamical spinodal calculation: the transition density is obtained from the crossing of the EoS at the temperature considered with the respective spinodal.
We can see that the 3DHFEOS and TF calculations give very compatible results, for the models here considered. When we compare with the thermodynamical spinodal calculation, we find that the difference of this method to the self-consistent calculations is $\sim 0.015$ fm$^{-3}$ for $T = 0$ MeV and decreases with increasing $T$.

\section{Conclusions}

The main conclusions of the present study are listed in the following. The effect of light clusters is weak and only noticeable at very low densities, but it was shown that taking clusters into account lowers the free energy. They will certainly affect transport properties. The density range of the pasta phase and the crust-core transition density decrease with increasing temperature, however the melting temperature of the different pasta phase geometries, or the inner crust, depend on the model properties. 
Our Skyrme models fall within the range of nuclear matter properties (especially the symmetry energy and its slope, see Table~\ref{tab1}) predicted by a number of experimental probes, and within this range, the pasta appearance and structure does not vary much. Both TF and 3DHFEOS calculations show jumps in the pressure that indicate a first order phase transition to uniform matter. Within the CP method, the inclusion of a non-homogeneous phase gives unrealistic results at low densities, though it predicts concordant transition densities to uniform matter. The stable geometries depend on the parametrizations used, and the properties that influence them should be investigated. It was also observed that $\beta$-equilibrium and fixed proton fraction matter show different melting temperatures for pasta, and different number of configurations. All the methods considered in this study show a very good agreement with respect to the transition density to homogeneous matter.

\subsection*{Acknowledgements}

Some of the RMF parametrizations data presented in this study, and referenced in the Figures, is taken from the collaboration work between the group from the University of Coimbra (S. Chiacchiera, F. Grill, C. Provid\^encia and I. Vida\~na) and the group from the Federal University of Santa Catarina, Florian\'opolis (S. S. Avancini and D. P. Menezes). 

H.P. is very grateful to the Organizers of the Workshop for the opportunity to present this work and for the financial support received. She is supported by FCT under Project No. SFRH/BPD/95566/2013. This work is partly supported by the project PEst-OE/FIS/UI0405/2014 developed under the initiative QREN. Partial support comes from ``NewCompStar'', COST Action MP1304.

\end{document}